\documentstyle[12pt,aps,epsf,floats]{revtex}

\begin{document}

\title{
\hspace*{\fill}\parbox[b]{5.5cm}{\small \tighten
hep-ph/0002032 \\
FERMILAB-Pub-00/032-T \\
CTEQ-015 \\
MSUHEP-00126 \\
UH-511-954-00 \\ }
\\
Higgs Production: \\ A Comparison of Parton Showers and Resummation}

\author{\\ 
C. Bal\'azs$^{1}$\footnote{balazs@phys.hawaii.edu}, 
J. Huston$^{2}$\footnote{huston@pa.msu.edu}, and 
I. Puljak$^{3}$\footnote{Ivica.Puljak@cern.ch}}

\address{
$^1$\,Department of Physics \& Astronomy, 
      University of Hawaii, Honolulu, HI 96822, USA \\
$^2$\,Department of Physics \& Astronomy, 
      Michigan State University, East Lansing, MI 48824, USA
$^3$\,Laboratoire de Physique Nucl\'eaire et des Hautes Energies, 
      Ecole Polytechnique, 91128 Palaiseau, France,
      and University of Split, 21000 Split, Croatia}

\date{August 29, 2000}

\maketitle
\thispagestyle{empty}

{\tighten
\begin{abstract}

The search for the Higgs boson(s) is one of the major priorities at the upgraded
Fermilab Tevatron and at the CERN Large Hadron Collider (LHC). Monte Carlo event
generators (MCs) are heavily utilized to extract and interpret the Higgs signal, 
which depends on the details of the soft--gluon emission from the initial state 
partons in hadronic collisions. Thus, it is crucial to establish the reliability 
of the MCs used by the experimentalists. In this paper, the MC based parton 
shower formalism is compared to that of an analytic resummation calculation. 
Theoretical input, predictions and, where they exist, data for the transverse 
momentum distribution of Higgs bosons, $Z^0$ bosons, and photon pairs are 
compared for the Tevatron and the LHC. This comparison is useful in 
understanding the strengths and the weaknesses of the different theoretical 
approaches, and in testing their reliability.

\end{abstract}
}

\draft



\section{Introduction}


To reveal the dynamics of the electroweak symmetry breaking, a new generation 
of hadron colliders will search for the Higgs boson(s). The potential 
of the upgraded Fermilab Tevatron, the 2 TeV center of mass energy 
proton-antiproton collider starting operation within a year, 
was analysed in Ref. \cite{PoR2W}. Later in this decade, two experimental 
collaborations (ATLAS and CMS) join the search at the CERN Large Hadron 
Collider (LHC) with 14 TeV proton-proton collisions. 
An extraction of the Higgs signal at the LHC requires not only the precise 
knowledge of the signal and background invariant mass distributions, but 
also the accurate prediction of the corresponding transverse momentum 
($p_T$) distributions. In general, the determination of the signal requires 
a detailed event modeling, an understanding of the detector resolution, 
kinematical acceptance and efficiency, all of which depend on the $p_T$ 
distribution. 
The shape of this distribution in the low to moderate $p_T$ region, 
can dictate the details of both the experimental triggering and the 
analysis strategies for the Higgs search.
It can also be used to devise an improved search strategy, and to enhance the 
statistical significance of the signal over the background 
\cite{Abdullin,BalazsNadolskySchmidtYuan}. 
In the $g g \to H X \to \gamma 
\gamma X$ mode at the LHC, for example, the shape of the signal and the 
background $p_T$ distribution of the photon pairs is different (c.f. Refs. 
\cite{BalazsYuanZZ,BalazsYuanShortH}), with the signal being harder. This 
difference can be utilized to increase the signal to background ratio. 
Furthermore, since vertex pointing with the photons is not possible in the 
CMS barrel, the shape of the $p_T$ distribution affects the precision of the 
determination of the event vertex from which the Higgs (decaying into two 
photons) originated.~\footnote{The vertex with the most activity is chosen as
the vertex from which the Higgs particle has originated. If the Higgs is 
typically produced at a relatively high value of $p_T$, then this choice is
correct a large fraction of the time.}
Thus, for a successful, high precision extraction of the 
Higgs signal, the theoretical calculation must be capable of reproducing 
the expected transverse momentum distribution. 


To reliably predict the $p_T$ distribution of Higgs bosons at the LHC, 
especially for the low to medium $p_T$ region where the bulk of the rate 
is, the effects of the multiple soft--gluon emission have to be included. 
One approach which achieves this is parton showering 
\cite{Sjostrand:1985xi}. Parton shower Monte Carlo programs such as {\tt 
PYTHIA}\cite{pythia}, {\tt HERWIG}\cite{herwig} and {\tt 
ISAJET}\cite{isajet} are commonly used by experimentalists, both as a way 
of comparing experimental data to theoretical predictions, and also as a 
means of simulating experimental signatures in kinematic regimes for which 
there are no experimental data yet (such as for the LHC). The final output 
of these Monte Carlo programs consists of the 4-momenta of a set of final 
state particles. This output can either be compared to reconstructed 
experimental quantities or, when coupled with a simulation of a detector 
response, can be directly compared to raw data taken by the experiment, 
and/or passed through the same reconstruction procedures as the raw data. 
In this way, the parton shower programs can be more useful to 
experimentalists than analytic calculations. Indeed, almost all of the 
physics plots in the ATLAS physics TDR~\cite{TDR} involve comparisons to 
{\tt PYTHIA} version 5.7. 


Predictions of the Higgs $p_T$ can also be obtained utilizing an analytic 
resummation formalism, which sums contributions of $\alpha_S^n 
\ln^m(m_H/p_T)$ (where $m_H$ is the Higgs mass, and $m \leq 2n-1$) up to all 
orders in the strong coupling $\alpha_S$. In the recent literature, most 
calculations of this kind are either based on, or originate from, the 
low $p_T$ factorization formalism%
\cite{CSS} (for the latest review see Ref. \cite{BCS}). This formalism 
resums the effects of the multiple soft--gluon emission while also 
systematically including the fixed order QCD corrections. 
It is possible to smoothly match the resummed result to the fixed 
order one in the intermediate to high $p_T$ region, thus obtaining a prediction 
for the full $p_T$ distribution \cite{BalazsYuanWZ}. In this paper, we use 
this formalism as the analytic `benchmark' to calculate the $p_T$ 
distributions of Higgs bosons at the LHC, and of $Z^0$ bosons and photon 
pairs produced in hadron collisions.


For many physical quantities, the predictions from parton shower Monte 
Carlo programs should be nearly as precise as those from analytic 
theoretical calculations. It is expected that both the Monte Carlo and 
analytic calculations should accurately describe the effects of the 
emission of multiple soft--gluons from the incoming partons, an all orders 
problem in QCD. The initial state soft--gluon emission affects the 
kinematics of the final state partons. This may have an impact on the 
signatures of physics processes at both the trigger and analysis levels 
and thus it is important to understand the reliability of such 
predictions. The best method for testing the reliability is a direct 
comparison of the predictions to experimental data. If no experimental 
data are available for certain predictions, then some understanding of the 
reliability may be gained from the comparison of the predictions from the 
two different methods. 

In the absence of experimental data for Higgs production, we can gauge the
reliability of calculations for this process by comparing them to each other.
We also compare predictions form the different formalisms to data for processes
which are similar to Higgs production at the LHC. In this way we can perform
a genuine `reality check' of the various theoretical predictions.
Production of a light, neutral Higgs boson at the LHC in the standard 
model (SM) and its supersymmetric extensions proceeds via the partonic 
subprocess $g g$ (through heavy fermion loop) $\to H X$. 
One of the major backgrounds for a light Higgs, in 
the mass range of 100  GeV $\lesssim m_H \lesssim$ 150 GeV, is diphoton 
production, a sizable contribution to which comes from the same, $g g$ 
initial state. Since the major part of the soft--gluon radiation is 
initiated from the incoming partons, the structures of the resummed 
corrections are similar for Higgs boson and diphoton production. Because 
the latter is measurable at the Fermilab Tevatron, diphoton production 
provides an exceptional opportunity to test the different theoretical 
models. $Z^0$ boson production can also be a good testing ground for the 
soft--gluon corrections to Higgs production. The treatment of the fixed 
order and resummed QCD corrections for $Z^0$ boson production is theoretically 
well understood and 
implemented at next-to-next-to-leading order \cite{BalazsYuanWZ}. 
Furthermore, just as in the diphoton case, predictions can also be 
tested against Tevatron data. The $Z^0$ data have the advantage that 
sufficient statistics exist in the Run 1 data from CDF and D0 to 
allow for detailed comparisons to the theoretical predictions.

\section{Low $p_T$ Factorization}


In this section the low transverse momentum factorization formalism and its 
matching to the usual factorization is reviewed. The problem arises as follows.
When calculating fixed order QCD corrections to the $p_T$ distribution of the
inclusive process $pp \to H X$, the standard QCD factorization theorem is 
invoked
\begin{equation} 
\frac{d\sigma}{dp_T^2} = 
f_{j_1/p}(m_H) \otimes
\frac{d{\hat \sigma}_{j_1j_2}}{dp_T^2} (m_H,p_T) \otimes
f_{j_2/p} (m_H),
\end{equation}
which is a convolution in the partonic momentum fractions, and is derived under
the usual assumption $p_T \gg m_H$.~\footnote{Here and henceforth, summation
on double partonic indices (e.g. $j_i$) is implied. 
Also, since we are focusing on the 
transverse momentum, longitudinal partonic momentum fractions are either kept 
implicit or, when applicable, integrated over.} When $p_T \ll m_H$ occurs, as 
a result of soft and soft+collinear emission of gluons from the initial state, 
the theorem fails. The ratio of the two very different physical scales in the 
partonic cross section ${\hat \sigma}_{j_1j_2}$, produces large logarithms of 
the form $\ln(m_H/p_T)$, which are not absorbed by the parton distribution 
functions $f_{j/p}$, unlike the ones originating from purely collinear parton 
emission. These logs are enhanced by a $1/p_T^2$ pre-factor at low $p_T$.
(The same factor suppresses them for large $p_T$.)
As a result, the Higgs $p_T$ distribution calculated using the 
conventional hadronic factorization theorem is un-physical in the low $p_T$ 
region.


To resolve this problem, the differential cross section is split into a part 
which contains all the logarithmic terms ($W$), and into a regular term ($Y$): 
\begin{equation}
\frac{d\sigma}{dp_T^2} = W(m_H,p_T) + Y(m_H,p_T) ,
\end{equation}
Since $Y$ does not contain logs of $p_T$, it can be calculated using the usual 
factorization. The $W$ term has to be evaluated differently, keeping in mind 
that failure of the standard factorization occurs because it neglects the 
transverse motion of the incoming partons in the hard scattering. As has been 
proven~\cite{CSS}, $W$ has a simple form in the Fourier conjugate, that is the 
transverse position (${\vec b}$) space
\begin{equation}
\widetilde{W}(m_H,b) = 
{\cal C}_{j_1/h_1}(m_H,b)\,
e^{-{\cal S}(m_H,b_*)}\,
{\cal C}_{j_2/h_2}(m_H,b) ,
\end{equation}
with the Sudakov exponent defined as
\begin{equation}
{\cal S}(m_H,b_*) = 
\int_{C_0^2/b_*^2}^{m_H^2} \frac{d \mu^2}{\mu^2}
\left[
A \left( \alpha_S(\mu) \right) \ln \left( \frac{m_H^2}{\mu^2} \right) + 
B \left( \alpha_S(\mu) \right)
\right] ,
\label{Eq:PerturbativeSudakov}
\end{equation}
which resums the large logarithmic terms.~\footnote{To prevent evaluation 
of the Sudakov exponent in the non-perturbative region, the impact 
parameter $b=|{\vec b}|$ is replaced by $b_* = b/\sqrt{1+(b/b_{\rm max })^2}$.
The choice of $C_0=2e^{- \gamma_E}$, where $\gamma_E$ is the Euler constant,
is customary.}
The partonic recoil against soft gluons, as well as the intrinsic partonic 
transverse momentum, are included in the generalized parton distributions
\begin{equation}
{\cal C}_{j/h}(m_H,b,x) =
\left[ C_{ja}\left( m_H,b_* \right) \otimes f_{a/h}(m_H) \right](x) \;
{\cal F}_{a/h}(m_H,b,x) ,
\label{CalC}
\end{equation}
where the convolution is evaluated over the partonic momentum fraction $x$.
The $A$ and $B$ functions, and the Wilson coefficients $C_{ja}$ are free 
of logs and safely calculable perturbatively as expansions in the strong
coupling
\begin{equation}
A(\alpha_S) =
\sum_{n=1}^\infty 
\left( \frac{\alpha _S}\pi \right)^n A^{(n)}, ~~~ {\rm etc.}
\end{equation}
The process independent non-perturbative functions ${\cal F}_{a/h}$, describing
long distance transverse physics, are extracted from low-energy 
experiments~\cite{Brock}.


The matching of the low and the high $p_T$ regions is achieved via the $Y$ 
piece. To correct the behavior of the resummed piece in the intermediate and 
high $p_T$ regions, it is defined as the difference 
of the cross section calculated by the standard factorization formula at a 
fixed order $n$ of perturbation theory and its $p_T\ll m_H$ 
asymptote.~\footnote{The expression for the $Y$ term 
for Higgs production can be found elsewhere~\cite{Yuan}.} The resummed cross 
section, to order $\alpha_S^n$, then reads as
\begin{equation}
\frac{d\sigma }{dp_T^2}=
W(m_H,p_T) +
      \frac{d\sigma^{(n)}}{dp_T^2} -
\left.\frac{d\sigma^{(n)}}{dp_T^2}\right|_{p_T\ll m_H}
\label{Eq:CSSMatched}
\end{equation}
At low $p_T$, the logarithms are large and the asymptotic part dominates the 
fixed order 
$p_T$ distribution. The last two terms in Eq.~(\ref{Eq:CSSMatched}) nearly 
cancel, and $W$ is a good approximation to the cross section. 
At high $p_T$ the logarithms are small, and the expansion of the resummed term
cancels the $p_T$ singular terms (up to higher orders in $\alpha_S$), 
and the cross section reduces to the fixed 
order perturbative result. After matching the resummed and fixed order cross 
sections in such a manner, it is expected that the normalization of the 
resummed cross section reproduces the fixed order total rate, since when 
expanded and integrated over $p_T$ it deviates from the fixed order result 
only in higher order terms.
For further details of the low $p_T$ factorization formalism and its 
application to Higgs production we refer to the recent 
literature~\cite{BalazsYuanShortH,BalazsMoriond}.

\section{Parton Showering and Resummation}

For technical reasons, the initial state parton shower proceeds by a 
$\it{backwards}$ evolution, starting at the large (negative) $Q^2$ 
scale of the hard scatter and then considering emissions 
at lower and lower (negative) virtualities, corresponding to earlier points on
the cascade (and earlier points in time), until a scale corresponding to the 
factorization scale is reached. The transverse momentum of 
the initial state is built up from the  whole series of splittings and 
boosts.
The showering process is independent of the hard scattering process being
considered (as long as one does not introduce any matrix element corrections),
and depends only on the initial state partons and the hard scale of
the process.

Parton showering utilizes the fact that the leading order 
singularities of cross sections factorize in the collinear limit.
This is expressed as
\begin{eqnarray}
   \lim_{p_b||p_g} |{\cal M}_{n+1}|^2 = 
   \frac{2 \pi \alpha_S}{p_b.p_g} P_{a\to b g}(z) |{\cal M}_{n}|^2, 
\end{eqnarray}
where ${\cal M}_{n+1}$ is the invariant amplitude for the process producing $n$ 
partons and a gluon, $\alpha_S$ is the strong coupling constant, $p_b$ and $p_g$ 
are the 4-momenta of the daughters of parton $a$, and $P_{a\to b g}(z)$ is the 
DGLAP evolution kernel associated with the $a \to b g$ splitting. These leading 
order collinear singularities can be factorized into a Sudakov form factor
\begin{eqnarray}
{\cal S}_{\rm shower}(Q) =  
\int_{Q_0^2}^{Q^2} \frac{d \mu^2}{\mu^2} \frac{\alpha_S(\mu)}{2\pi} 
\int_{0}^{1} dz \, P_{a\to bg}(z),
\end{eqnarray}
which is interpreted as the probability ${\cal P}=\exp(-{\cal S}_{\rm shower})$
of the partonic evolution from scale $Q_0$ to $Q$ with no resolvable branchings.
This probability
can be used to determine the scale for the first emission and 
hence for the whole cascade. The formalism can be extended to soft singularities 
as well by using angular ordering. In this approach, the choice of the hard 
scattering is based on the use of evolved parton distributions, which means that 
the inclusive effects of initial-state radiation are already included. What 
remains is, therefore, to construct the exclusive showers.

Parton showering resums primarily the leading logs which are universal, 
that is process independent and depend only on the given initial state. 
In this lies one of the strengths of Monte Carlos, since parton showering can
be incorporated into a wide variety of physical processes.
An analytic calculation, in comparison, can resum all logs. For example, the 
low $p_T$ factorization
formalism sums all of the logarithms with $m_H/p_T$ in their arguments.
As discussed earlier, all of the `dangerous logs' are included in the 
Sudakov exponent (\ref{Eq:PerturbativeSudakov}).
The $A$ and $B$ functions in Eq.(\ref{Eq:PerturbativeSudakov})
contain an infinite number of coefficients, 
with the $A^{(n)}$ coefficients being universal, while the $B^{(n)}$'s are
process dependent, with the exception of $B^{(1)}$. 
In practice, the number of towers of logarithms included in the analytic Sudakov 
exponent
depends on the level to which a fixed order calculation was performed for a 
given process. 
Generally, if a next-to-next-to-leading order calculation is available, then 
$B^{(2)}$ can be extracted and incorporated. 
Extraction of higher coefficients require the knowledge of even higher order 
QCD corrections. So far, only the $A^{(1)}$,
$A^{(2)}$ and $B^{(1)}$ coefficients are known for Higgs production but the 
calculation of $B^{(2)}$ is in progress~\cite{carlschmidt}.
If we try to interpret parton showering in the same language
then we can say that the Monte Carlo Sudakov exponent always
contains a term analogous to $A^{(1)}$. 
It was shown in Ref.~\cite{webber} that a term equivalent to $B^{(1)}$ is also
included in the ({\tt HERWIG}) shower algorithm, and a suitable modification of
the Altarelli-Parisi splitting function, or equivalently the strong coupling 
constant $\alpha_S$, also effectively approximates the $A^{(2)}$ 
coefficient.~\footnote{This is rigorously true only for the high x or 
$\sqrt{\tau}$ region.}

In contrast with the shower Monte Carlos, analytic resummation 
calculations integrate over the kinematics of the soft--gluon emission, with the
result that they are limited in their predictive power to
inclusive final states. While the Monte Carlo maintains an exact treatment 
of the branching kinematics, in the original low $p_T$ factorization 
formalism no kinematic penalty is paid for the emission of the soft--gluons, 
although an approximate treatment of this can be incorporated into 
its numerical implementations, such as 
ResBos~\cite{BalazsYuanShortH,BalazsYuanWZ}.
Neither the parton showering process nor the analytic resummation
translate smoothly into kinematic configurations
where one hard parton is emitted at large $p_T$. In the Monte Carlo matrix 
element corrections, and in the analytic resummation calculation matching,
is necessary. This matching is standard procedure for resummed calculations, 
and matrix element corrections are becoming increasingly common in 
Monte Carlos~\cite{pythiacor,herwigcor,isajetcor}.
 
With the appropriate input from higher order cross sections,
a resummation calculation has the corresponding higher order normalization and
scale dependence. 
The normalization and
scale dependence for the Monte Carlo, though, remains that of a leading 
order calculation. The parton showering process redistributes the events 
in phase space, but does not change the total cross section (for
example, for the production of a Higgs boson).~\footnote{%
Technically, one could add the branching for $q\to q$+Higgs in the shower, 
which would somewhat increase the Higgs cross section. However, the main 
contribution to the higher order $K$-factor comes from the virtual 
corrections and the `Higgs bremsstrahlung' contribution is negligible.}

One quantity which is expected to be well described by both 
calculations is the transverse momentum
of the final state electroweak boson in a subprocess such
as $q\overline{q} \to W^\pm X$, $Z^0X$ or $gg \to H X$, where most of the 
$p_T$ is provided by initial state parton emission. The parton showering 
supplies the same sort of transverse kick as the resummed soft--gluon emission
in the analytic calculation. Indeed, similar 
Sudakov form factors appear in both approaches.
The correspondence between the Sudakov form factors of
resummation and Monte Carlo approaches embodies many 
subtleties, relating to both the arguments of the
Sudakov factors as well as the impact of sub-leading logs~\cite{mrenna}.

At a point in its evolution, typically corresponding to the virtuality of a 
few GeV$^2$, the parton shower is cut off and the effects of gluon emission at
softer scales must be parameterized and inserted by hand.  This is similar to
the somewhat  arbitrary division between perturbative and non-perturbative
regions in the resummation calculation. The parameterization is typically
expressed in a Gaussian form, similar to that used for the non-perturbative 
$k_T$ in a resummation program~\cite{Brock}. In general, the value for the 
non-perturbative $\langle k_T \rangle$ needed in a Monte Carlo program will 
depend on the particular kinematics and initial state being investigated. 
	A value of the average non-perturbative 
$k_T$ of greater than 1 GeV, for example, does not imply that there
is an anomalous intrinsic $k_T$ associated with the parton size. Rather, this
amount of $\langle k_T \rangle$ needs to be supplied to provide what is 
missing in the 
truncated parton shower. If the shower is cut off at a higher virtuality, more
of the `non-perturbative' $k_T$ will be needed.

\section{$Z^0$ Boson Production at the Tevatron}

From a theoretical viewpoint, $Z^0$ production at the Tevatron is one of the 
highest precision testing grounds for the effects of multiple soft-gluon emission.
The fully differential fixed order cross section has been calculated up to 
${\cal O}(\alpha_S^2)$, 
and the $A^{(1,2)}$, $B^{(1,2)}$, and $C^{(1)}$ resummed coefficients are 
known for this process, and have been numerically implemented 
\cite{BalazsYuanWZ}. Since the ${\cal O}(\alpha_S^2)$ corrections are 
relatively small (the order of a percent), the contribution of $B^{(2)}$ 
is almost negligible. Thus, nominally the same perturbative physics is 
implemented in the shower Monte Carlos as in the resummation calculation. 
Any differences between their predictions can be ascribed to the small
differences in the implementation of the perturbative physics, and to the 
different non-perturbative physics they contain. Experimentally, the 
4-momentum of a $Z^0$ boson, and thus its $p_T$, 
can be measured with great precision in the $e^+e^-$ decay mode. 
Resolution effects are relatively minor and are easily corrected. Thus, 
the $Z^0$ $p_T$ distribution is an excellent probe of the effects of the 
soft--gluon emission. 

\begin{figure}[t]
\begin{center}
\epsfxsize=12cm
\epsfysize=10cm
\mbox{\epsfbox{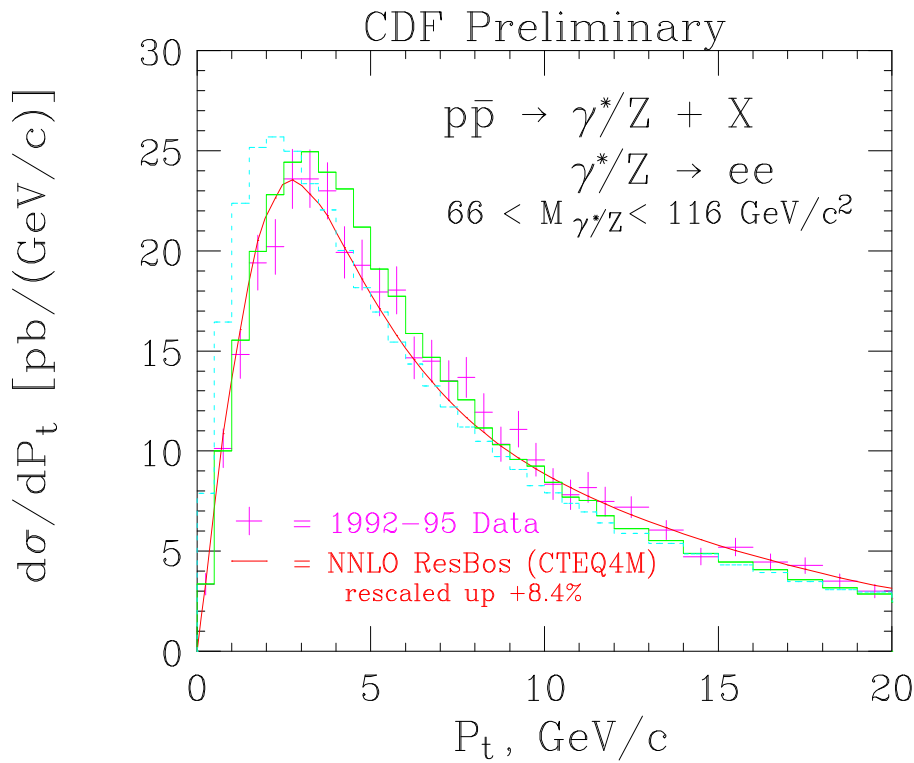}}
\end{center}
\caption{ \sf 
The $Z^0$ $p_T$ distribution (at low $p_T$) from CDF for Run 1 compared to 
predictions from ResBos (curve) and from {\tt PYTHIA} (histograms). The 
two {\tt PYTHIA} predictions use the default (rms) value for the non-%
perturbative $k_T$ (0.44 GeV) and the value that gives the best agreement 
with the shape of the data (2.15 GeV). The normalization of the resummed 
prediction was rescaled upwards by 8.4\%. The {\tt PYTHIA} prediction 
was rescaled by a factor of 1.4 for the shape comparison.
(Including only soft--gluon QCD corrections, {\tt PYTHIA} does not contain the 
QCD $K$-factor.)
}
\label{fig:run1_ee_pt}
\end{figure}

The resolution corrected $p_T$ distribution (in the low $p_T$ region) 
for $Z^0$ bosons from the CDF experiment 
\cite{Affolder:1999jh} is shown in Figure~\ref{fig:run1_ee_pt}, compared 
to both the resummed prediction of ResBos, and to two predictions from 
{\tt PYTHIA} version 6.125.
One {\tt PYTHIA} prediction uses the default value of intrinsic $k_T^{\rm rms} 
= 0.44$ GeV (dashed histogram)\footnote{For a Gaussian distribution, 
$k_T^{\rm rms}=1.13\langle k_T \rangle$.}, and the second a value of 2.15 GeV 
(solid histogram), per incoming parton.~\footnote{A previous 
publication~\cite{pythiacor} indicated the need for a substantially larger 
non-perturbative $\langle k_T \rangle$, of the order of 4 GeV, for the 
case of $W^\pm$ production at the Tevatron. The data used in the comparison, 
however, were not corrected for resolution smearing, a fairly large effect 
for the case of $W \to e{\nu}$ production and decay.} The latter 
value was found to give the best agreement for {\tt PYTHIA} with the 
data, and a similar conclusion has been reached in comparisons of 
the CDF $Z^0$ $p_T$ data with {\tt HERWIG}~\cite{corcella}.
All of the predictions use the CTEQ4M parton distributions~\cite{cteq4}.
The shift between 
the two {\tt PYTHIA} predictions at low $p_T$ is clearly evident. As might 
have been expected, the high $p_T$ region (above 10 GeV) is unaffected by the 
value of the non-perturbative $k_T$. Much of the $k_T$  `given' 
to the incoming partons at their lowest virtuality, $Q_0$, is reduced at 
the hard scatter due to the number of gluon branchings preceding the 
collision. The emitted gluons carry off a sizable fraction of the 
original non-perturbative $k_T$~\cite{sjostrandpc}. 
This point will be investigated in more 
detail later for the case of Higgs production. 

In the resummed calculation it has been shown that, in addition to the 
perturbative physics (Sudakov and Wilson coefficients, $C_{ja}$), the choice 
of the non-perturbative parameters affects the shape of the distribution in 
the lowest $p_T$ region and the location of the 
peak~\cite{BalazsYuanWZ}. In order to qualitatively compare 
this effect to the smearing applied in the Monte Carlos,
it is possible to bring the resummation formula to a form where the 
non-perturbative function acts as a Gaussian type smearing term. 
This, using the Ladinsky-Yuan parameterization~\cite{LY} of 
the non-perturbative function, leads to an rms value of 2.5 GeV for the 
effective $k_T$ smearing parameter, for $Z^0$ production at the Tevatron. 
This is in agreement with the {\tt PYTHIA} and {\tt HERWIG} results:
to well describe the $Z^0$ production data at the Tevatron, 2-2.5 GeV
non-perturbative $k_T$ is needed in these implementations. 
%
%
The resummed curve agrees with the shape of the data well, which is a 
non-trivial result, since the resummation calculation does not contain any 
free parameters which are fitted to the $Z^0$ $p_T$ distribution.
Even with the 
optimal non-perturbative $k_T^{\rm rms} = 2.15$ GeV, there is slight shape 
difference between the shower Monte Carlo and the data.
This might be partially due to the lack of the $B^{(2)}$ coefficient in
the shower Monte Carlos. 
This is supported by the fact that,
if the $B^{(2)}$ coefficient was not included in the resummed prediction, 
the result would be an increase in the height of the peak and a decrease in 
the rate between 10 and 20 GeV, leading to a better agreement with the
best {\tt PYTHIA} prediction~\cite{BCS}.

\begin{figure}[t]
\begin{center}
\epsfxsize=12cm
\epsfysize=10cm
\mbox{\epsfbox{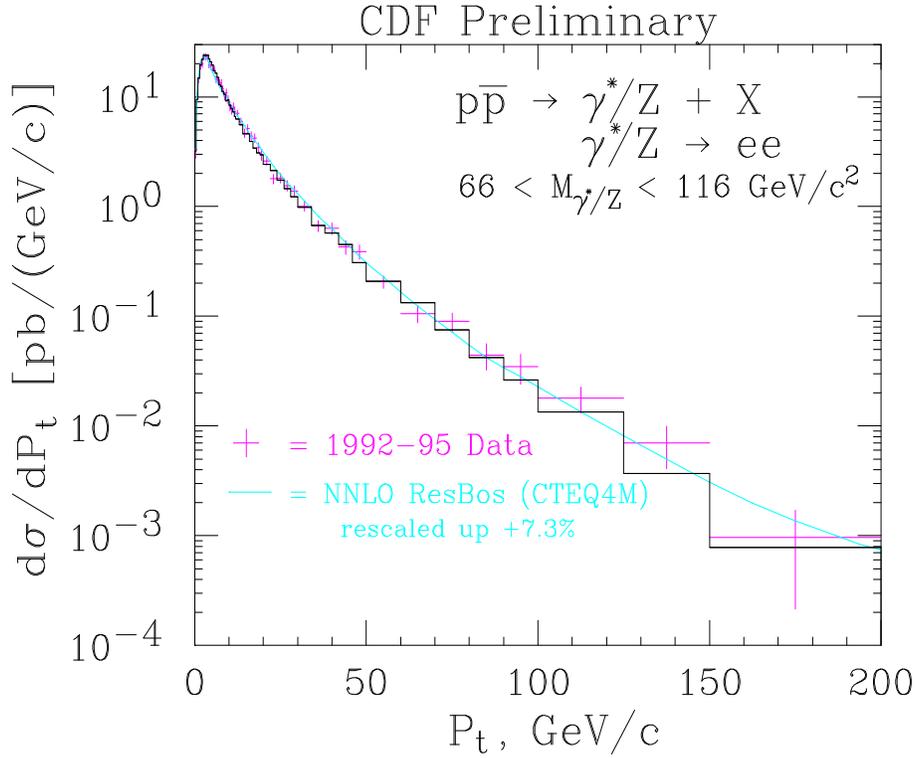}}
\end{center}
\caption{
\sf The $Z^0$ $p_T$ distribution (for the full range of $p_T$) from CDF for 
Run 1 compared to predictions from ResBos (curve) and from {\tt PYTHIA}  
(histogram). The normalization of the resummed 
prediction was rescaled upwards by 8.4\%. The {\tt PYTHIA} prediction 
was rescaled by a factor of 1.4 for the shape comparison.
} 
\label{fig:zpteeall}
\end{figure}
The $Z^0$ $p_T$ distribution is shown over a wider $p_T$ range in  
Figure~\ref{fig:zpteeall}. The {\tt PYTHIA} and ResBos predictions both 
describe the data well. Note especially the agreement of {\tt PYTHIA} with 
the data at high $p_T$, made possible by explicit matrix element 
corrections (from the subprocesses $q\overline{q} \to Z^0g$ and 
$gq \to Z^0q$) to the $Z^0$ production process.~\footnote{Slightly 
different techniques are used for the matrix element corrections by {\tt 
PYTHIA}~\cite{pythiacor} and by {\tt HERWIG}~\cite{herwigcor}.  In {\tt 
PYTHIA}, the parton shower probability distribution is applied over the 
whole phase space and the exact matrix element corrections are applied 
only to the branching closest to the hard scatter.  In {\tt HERWIG}, the 
corrections are generated separately for the regions of phase space 
unpopulated by {\tt HERWIG} (the `dead zone') and the populated region. In 
the dead zone, the radiation is generated according to a distribution 
using the first order matrix element calculation, while the algorithm for 
the already populated region applies matrix element corrections whenever a 
branching is capable of being `the hardest so far'.}

\section{Diphoton Production}

\begin{figure}[t]
\begin{center}
\epsfxsize=12cm
\epsfysize=12cm
\vspace{-2cm}
\mbox{ \epsfbox{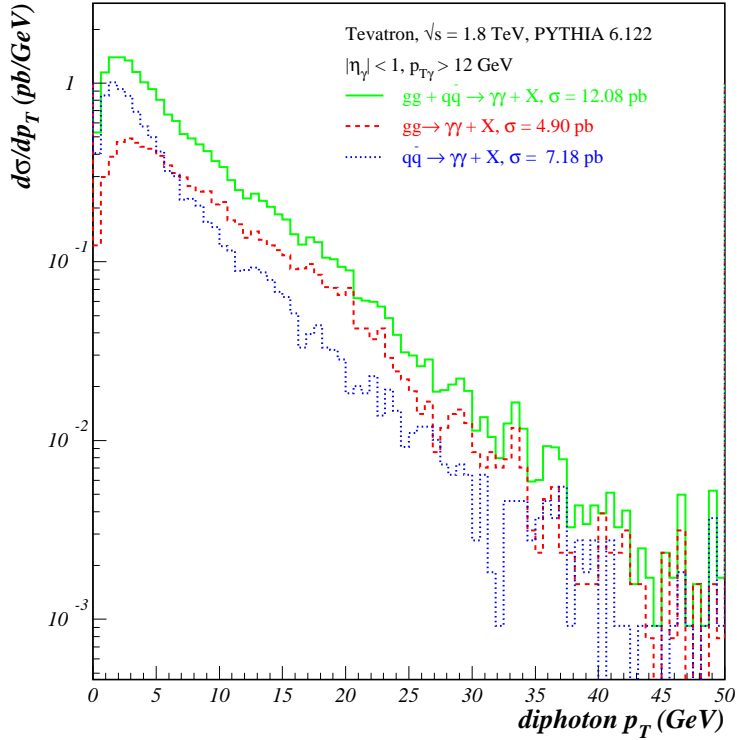}}
\end{center}
\caption{
\sf A comparison of the {\tt PYTHIA} predictions for diphoton production at 
the Tevatron for the two different subprocesses, $q\overline{q}$ and $gg$. 
The same cuts are applied to {\tt PYTHIA} as in the CDF diphoton analysis.
} 
\label{fig:pythiakt}
\end{figure}

Most of the experience that we have for comparisons of data to resummation 
calculations or Monte Carlos is based on Drell-Yan pair production, that 
is mostly on $q\overline{q}$ initial states. It is important then to 
examine diphoton production at the Tevatron, where a large fraction of the 
contribution at low mass is due to $gg$ scattering. The prediction for the 
diphoton $p_T$ distribution at the Tevatron, from {\tt PYTHIA} (version 
6.122), is shown in Figure~\ref{fig:pythiakt}, using the experimental cuts 
applied in the CDF analysis~\cite{cdfdiphot}. About half of the diphoton 
cross section at the Tevatron is due to the $gg$ subprocess, and the 
diphoton $p_T$ distribution is noticeably broader for the $gg$ subprocess 
than for the $q\overline{q}$ subprocess.

\begin{figure}[t]
\begin{center}
\epsfxsize=12cm
\epsfysize=12cm
\vspace{-2cm}
\mbox{\epsfbox{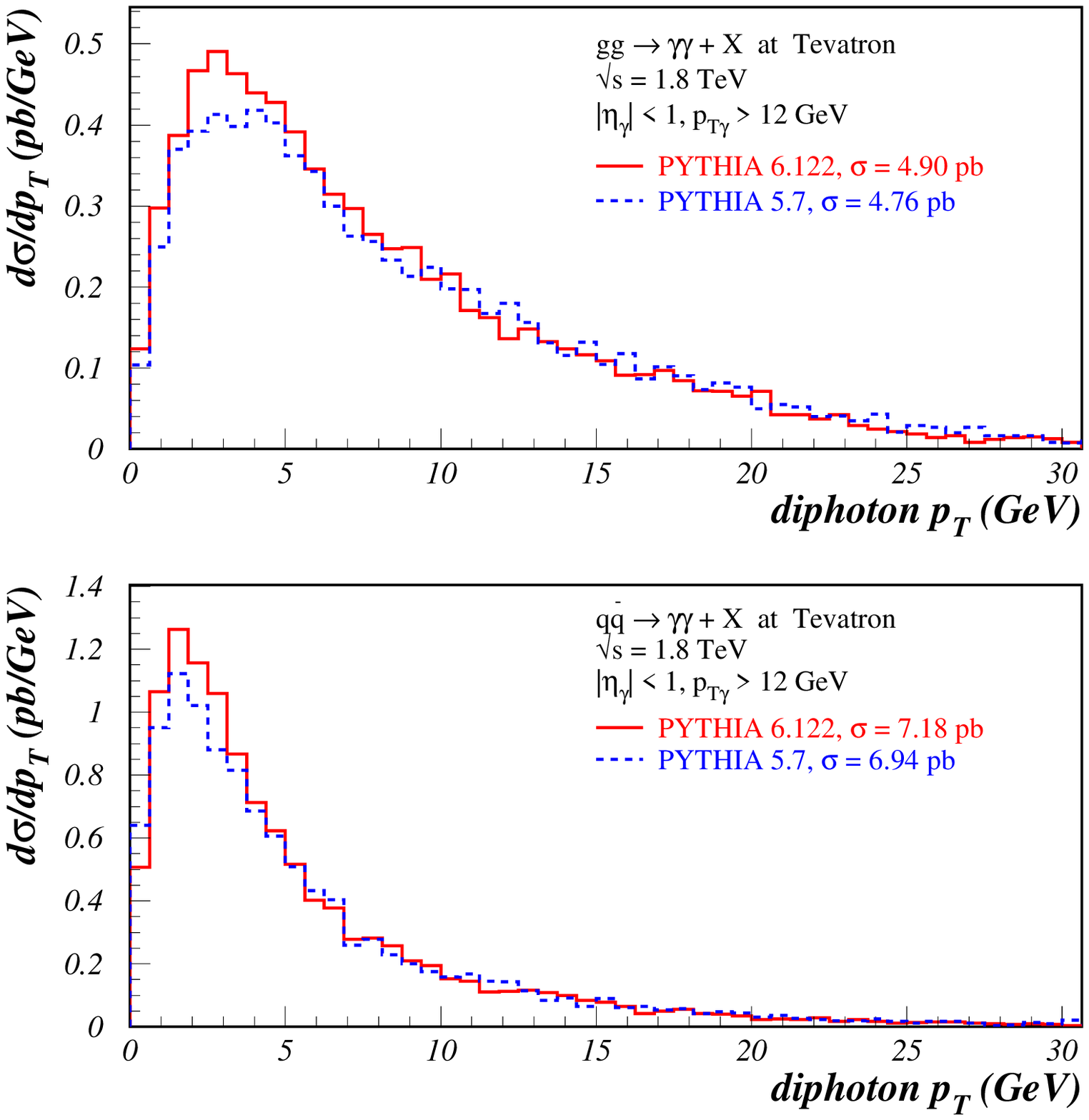}}
\end{center}
\caption{
\sf A comparison of the {\tt PYTHIA} predictions for diphoton production 
at the Tevatron for the two different subprocesses, $gg$ (top) and 
$q\overline{q}$ (bottom), for two recent versions of {\tt PYTHIA}. The 
same cuts are applied to {\tt PYTHIA} as in the CDF diphoton analysis. 
} 
\label{fig:2gamma_tev}
\end{figure}

\begin{figure}[t]
\begin{center}
\epsfxsize=12cm
\epsfysize=12cm
\vspace{-2cm}
\mbox{\epsfbox{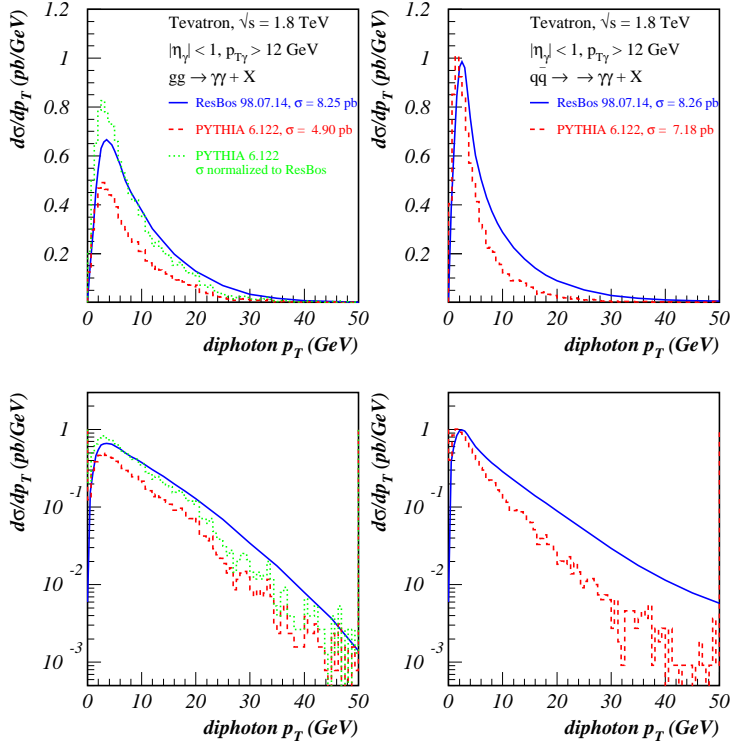}}
\end{center}
\caption{
\sf A comparison of the {\tt PYTHIA} and ResBos predictions for diphoton 
production at the Tevatron for the two different subprocesses, $gg$ (left)
and $q\overline{q}$ (right). The same cuts are applied to {\tt PYTHIA} and 
ResBos as in the CDF diphoton analysis.
The bottom figures show the same in logarithmic scale.
} 
\label{fig:pythrestev}
\end{figure}

A comparison of the $p_T$ distributions for the two diphoton subprocesses 
$(q\overline{q}, gg)$ in two recent versions of {\tt PYTHIA}, 5.7 and 6.1, 
is shown in Figure~\ref{fig:2gamma_tev}. There seems to be
little difference in the 
 $p_T$ distributions between the two versions for both subprocesses. As will
be shown later, this is not true for the case of Higgs production. 
%
In  Figure~\ref{fig:pythrestev} are shown the ResBos predictions for 
diphoton production at the Tevatron from $q\overline{q}$ and $gg$ 
scattering compared to the {\tt PYTHIA} predictions. 
The $gg$ subprocess predictions in ResBos agree well 
with those from {\tt PYTHIA} 
while the $q\overline{q}$ $p_T$ distribution is noticeably broader in 
ResBos. The latter behavior is due to the presence of the $Y$ piece in 
ResBos at moderate $p_T$, that is the matching of the  $q\overline{q}$ cross 
section to the fixed order $q\overline{q} \to {\gamma}{\gamma}g$ at high $p_T$.
The corresponding matrix element correction is not implemented in {\tt PYTHIA}. 
The {\tt PYTHIA} and ResBos predictions for $gg 
\to {\gamma}{\gamma}$ agree in the moderate $p_T$ region, even though the 
ResBos prediction has the $Y$ piece present and is matched to the matrix 
element piece $gg \to {\gamma}{\gamma}g$ at high $p_T$, while there is no 
such matrix element correction for {\tt PYTHIA}. This demonstrates the smallness
of the $Y$ piece for the $gg$ subprocess, which is the same conclusion that was 
reached in Ref.~\cite{BalazsNadolskySchmidtYuan}.
One way to understand this is recalling that the $gg$ parton-parton 
luminosity falls very steeply with increasing partonic center of mass energy, 
$\sqrt{\hat{s}}$. This falloff 
tends to suppress the size of the $Y$ piece since the production of the 
diphoton pair at higher $p_T$ requires larger values of longitudinal momentum
fractions. 

\begin{figure}[t]
\begin{center}
\epsfxsize=12cm
\epsfysize=12cm
\vspace{-2cm}
\mbox{\epsfbox{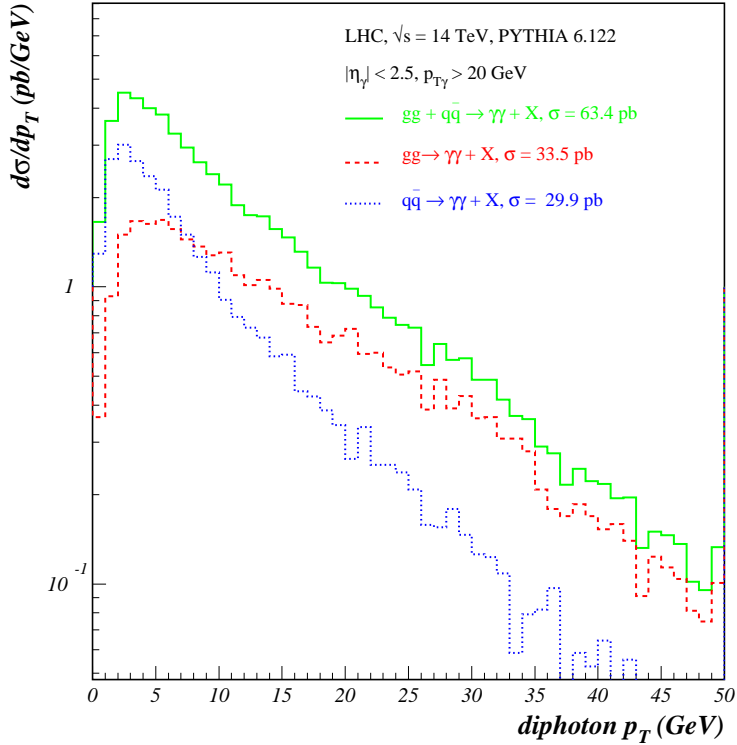}}
\end{center}
\caption{
\sf A comparison of the {\tt PYTHIA} predictions for diphoton production 
at the LHC for the two different subprocesses, $q\overline{q}$ and $gg$. 
Similar cuts are applied to the diphoton kinematics as those used by ATLAS 
and CMS.} 
\label{fig:lhc_diphot}
\end{figure}

Comparisons of the diphoton data measured by both the CDF~\cite{cdfdiphot} 
and D0~\cite{d0diphot} experiments indicate a disagreement of the observed 
diphoton $p_T$ distribution with the NLO QCD predictions~\cite{owens}. In 
particular, the $p_T$ distribution in the data is noticeably broader than 
that predicted by fixed order QCD calculations, but in agreement with the 
predictions of ResBos \cite{BalazsBergerMrennaYuan}. 
The transverse distributions of the 
diphoton pair are particularly sensitive to the effects of soft--gluon 
radiation. The $p_T$ distribution, for example, is a delta 
function calculated at leading order, and is strongly smeared by the 
all order Sudakov factor. 
Given the small size of the diphoton cross section at the Tevatron, the 
comparisons for Run 1 are statistically limited. A more precise 
comparison between theory and experiment will be possible with 
the 2 $fb^{-1}$ or greater data sample that is expected for CDF and D0 in 
Run 2, and at the LHC. 
The Monte Carlo prediction for the diphoton production cross section, as a 
function
of the diphoton $p_T$ and using cuts appropriate to ATLAS and CMS, is shown
in Figure~\ref{fig:lhc_diphot}. As at the Tevatron, about
half of the cross section is due to $gg$ scattering and the diphoton 
$p_T$ distribution from $gg$ scattering is noticeably broader than that from
$q\overline{q}$ production. 

\begin{figure}[t]
\begin{center}
\epsfxsize=12cm
\epsfysize=12cm
\vspace{-2cm}
\mbox{\epsfbox{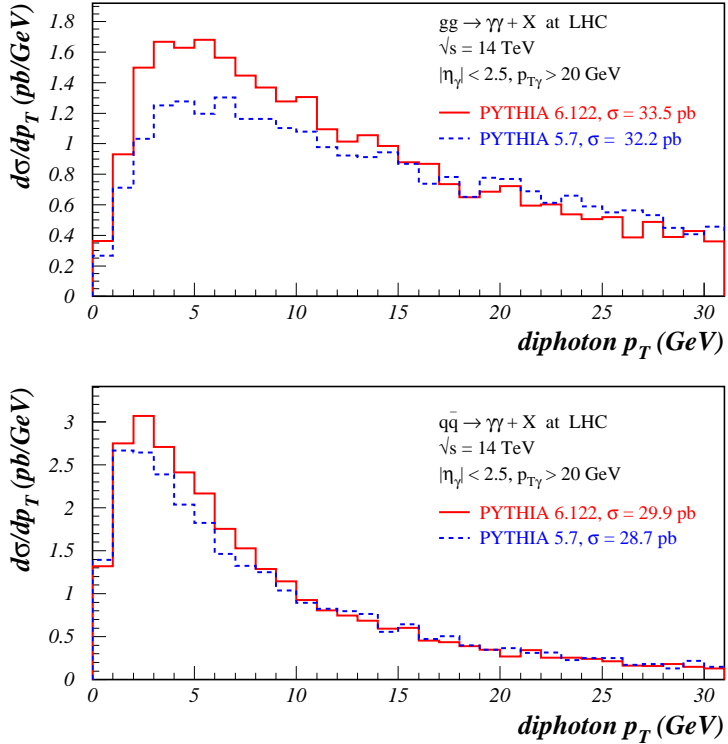}}
\end{center}
\caption{
\sf A comparison of the {\tt PYTHIA} predictions for diphoton production 
at the LHC for the two different subprocesses, $gg$ (top) and 
$q\overline{q}$ (bottom), for two recent versions of {\tt PYTHIA}. Similar 
cuts are applied to the diphoton kinematics as are used by ATLAS and CMS. 
} 
\label{fig:lhc_diphot_57}
\end{figure}

\begin{figure}[t]
\begin{center}
\epsfxsize=12cm
\epsfysize=12cm
\vspace{-2cm}
\mbox{\epsfbox{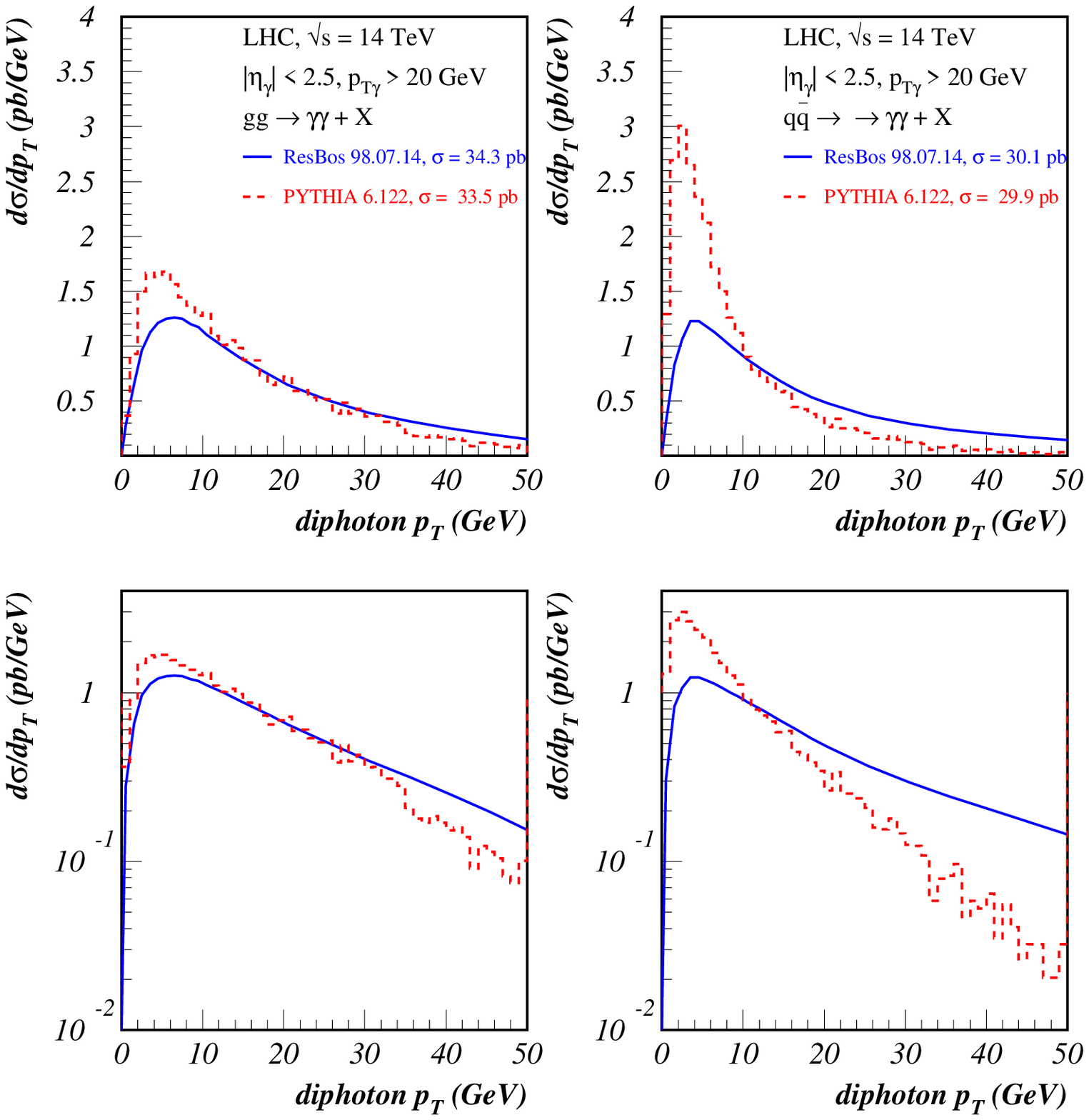}}
\end{center}
\caption{
\sf A comparison of the {\tt PYTHIA} and ResBos predictions for diphoton 
production at the LHC for the two different subprocesses, $gg$ (left) and 
$q\overline{q}$ (right). Similar cuts are applied to {\tt PYTHIA} and 
ResBos as in the ATLAS and CMS diphoton analyses. The bottom figures show 
the same in logarithmic scale.
} 
\label{fig:pythreslhc}
\end{figure}

In Figure~\ref{fig:lhc_diphot_57} is shown a comparison of the diphoton 
$p_T$ distribution at the LHC for two different versions of {\tt PYTHIA}, 
for the two 
different subprocesses. Note that the $p_T$ distribution in {\tt PYTHIA} 
version 5.7 is somewhat broader than that in version 6.122 for the case of 
$gg$ scattering. The effective diphoton mass range being considered here 
is lower than the 150 GeV Higgs mass that will be considered in the next 
section. As will be seen, the differences in soft--gluon emission between 
the two versions of {\tt PYTHIA} are larger in that case. 
In Figure~\ref{fig:pythreslhc} are shown the ResBos predictions for 
diphoton production at the LHC from $q\overline{q}$ and $gg$ scattering 
compared to the {\tt PYTHIA} predictions. 
Again, the $gg$ subprocess in ResBos agree well with 
{\tt PYTHIA}, while the $q\overline{q}$ $p_T$ distribution is 
noticeably broader in ResBos, for the reasons cited previously. 

\section{Higgs Boson Production}

\begin{figure}[t]
\begin{center}
\epsfxsize=12cm
\epsfysize=12cm
\vspace{-2cm}
\mbox{\epsfbox{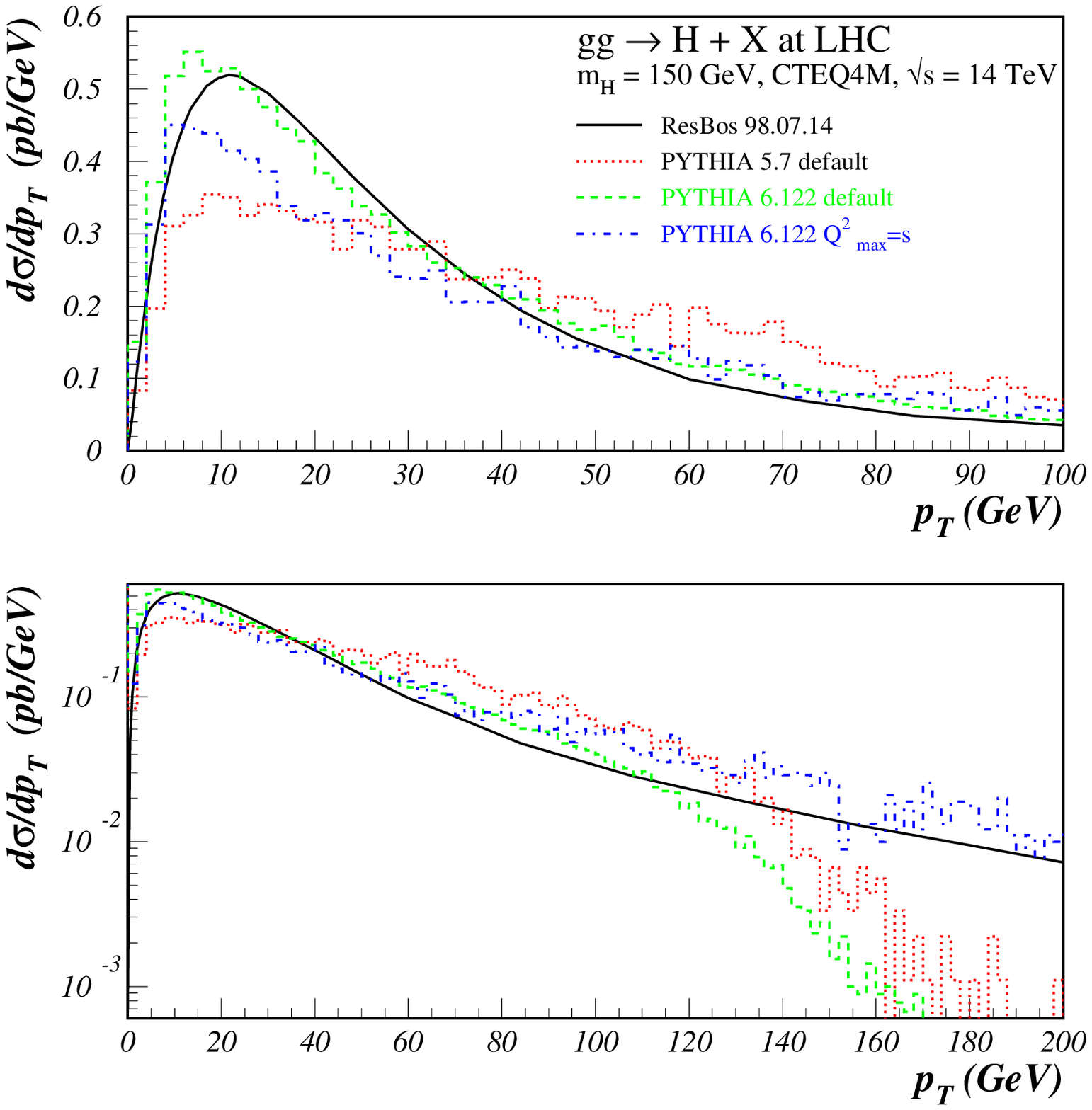}}
\end{center}
\caption{
\sf A comparison of predictions for the Higgs $p_T$ distribution at the LHC 
from ResBos and from two recent versions of {\tt PYTHIA}. The ResBos and 
{\tt PYTHIA} predictions have been normalized to the same area. 
The bottom figure shows the same in logarithmic scale.
} 
\label{fig:resbos_pythia}
\end{figure}

A comparison of the SM Higgs $p_T$ distribution at the LHC, for a Higgs mass of 
150 GeV, is shown in Figure~\ref{fig:resbos_pythia}, for ResBos and the two 
recent versions of {\tt PYTHIA}. As before, {\tt PYTHIA} has been rescaled
by a factor of 1.7,
to agree with the normalization of ResBos to allow for a better shape 
comparison. There are a number of features of interest. 
First, the peak of the resummed distribution has moved to $p_T 
\approx$ 11 GeV (compared to about 3 GeV for $Z^0$ production at the Tevatron). 
This is  partially due to  the larger mass (150 GeV compared to 90 GeV), 
but is primarily because of the larger color factors associated with 
initial state gluons ($C_A = 3$) rather than quarks ($C_F = 4/3$), and also 
because of the larger phase space for initial state gluon emission at the LHC.
Second, and more importantly, there is a substantial disagreement for the shape 
of the Higgs $p_T$ distribution between ResBos and {\tt PYTHIA} 5.7, and 
between the two versions of {\tt PYTHIA}. An understanding of the reasons for 
these differences is critical, as the shape of the transverse momentum 
distribution for the Higgs in the low to moderate $p_T$ region, can dictate the
details of both the experimental triggering and the analysis strategies for the
Higgs search. As noted before,  most of the studies for Higgs production by
CMS and ATLAS have been based on {\tt PYTHIA} 5.7.
For the CMS detector, the higher $p_T$ 
activity associated with Higgs production in version 5.7  
allows for a more precise determination of the event vertex from which 
the Higgs (decaying into two photons) originates. Vertex pointing with the 
photons is not possible in the CMS barrel, and the large number of interactions 
occurring with high intensity running will mean a substantial probability 
that  at least one of the interactions will
have more activity than the Higgs vertex, thus leading to the assignment of 
the Higgs decay to the wrong vertex, and therefore a noticeable degradation 
of the $\gamma \gamma$ effective mass resolution.

In comparison to ResBos, the older version of {\tt PYTHIA} produces too 
many Higgs events at moderate $p_T$. 
Two changes have been implemented in the newer version 6.1. The first change 
is that a cut is placed on the combination of $z$ 
(longitudinal momentum fraction) and $Q^2$ (partonic virtuality) values in a 
branching: $\hat{u} = Q^2-\hat{s}(1-z) < 0$, where $\hat{s}$ refers to the 
squared invariant mass of the subsystem of the hard scattering plus the shower 
partons considered to 
that point.  The association with $\hat{u}$ is relevant if the branching 
is interpreted in terms of a $2 \to 2$ hard scattering. The corner of 
emissions that do not respect this requirement occurs when the $Q^2$ value 
of the space-like emitting parton is little changed and the $z$ value of 
the branching is close to unity. This effect is mainly for the hardest 
emission (largest $Q^2$). The net result of this requirement is a 
substantial reduction in the total amount of gluon radiation 
\cite{pythiaman}.~\footnote{Such branchings are kinematically allowed, but 
since matrix element corrections would assume initial state partons to 
have $Q^2=0$, a non-physical $\hat{u}$ results (and thus  no possibility 
to impose matrix element corrections). The correct behavior is beyond the 
predictive power of leading log Monte Carlos.} In the second change, the 
parameter for the minimum gluon energy emitted in space-like showers is 
modified by an extra factor roughly corresponding to the $1/\gamma$ factor 
for the boost to the hard subprocess frame~\cite{pythiaman}. The effect of 
this change is to increase the amount of gluon radiation. Thus, the two 
effects are in opposite directions but with the first effect being 
dominant. In principle, this problem could affect the $p_T$ distribution for all 
{\tt PYTHIA} processes. In practice, it affects only $gg$ initial states, 
due to the enhanced probability for branching in such an initial state. 

The newer version of {\tt PYTHIA} agrees well with ResBos at low to 
moderate $p_T$, but falls below the resummed prediction at high $p_T$.
The agreement of the predictions of {\tt PYTHIA} 6.1 with those of ResBos,
in the low to moderate $p_T$ region, gives some credence that the changes
made in {\tt PYTHIA} move in the right direction.  
The disagreement at high $p_T$ can be easily understood: 
ResBos switches to the NLO Higgs+jet matrix 
element at high $p_T$, while the default {\tt PYTHIA} can generate the 
Higgs $p_T$ distribution only by initial state gluon radiation, using 
the Higgs mass squared as maximum virtuality. High $p_T$ Higgs production is 
another example where a $2 \to 1$ Monte Carlo calculation with parton 
showering cannot completely reproduce the exact matrix element 
calculation, without the use of matrix element corrections. The high $p_T$ 
region is better reproduced if the maximum virtuality $Q_{max}^2$ is set 
equal to the squared partonic center of mass energy, $s$, rather than 
$m_H^2$. This is equivalent to applying the parton shower to all of phase 
space. However, this has the consequence of depleting the low $p_T$ region, 
as `too much' showering causes  events to migrate out of the peak.  The 
appropriate scale to use in {\tt PYTHIA} (or any Monte Carlo) depends on 
the $p_T$ range to be probed.  If matrix element information is used to 
constrain the behavior, the correct high $p_T$ cross section can be 
obtained while still using the lower scale for showering. Thus, the 
incorporation of matrix element corrections to Higgs production (involving 
the processes $gq \to qH$,$q{\overline{q}} \to gH$, $gg \to gH$) is the 
next logical project for the Monte Carlo experts, in order to accurately 
describe the high $p_T$ region, and is already in 
progress~\cite{sjostrandpc,corcellapc}.

\begin{figure}[t]
\begin{center}
\epsfxsize=12cm
\epsfysize=12cm
\vspace{-2cm}
\mbox{\epsfbox{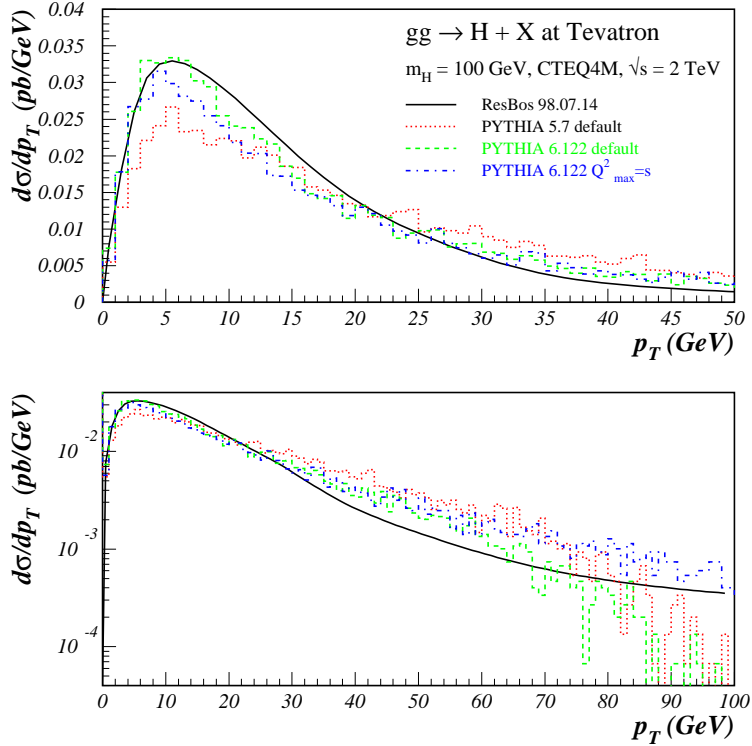}}
\end{center}
\caption{
\sf A comparison of predictions for the Higgs $p_T$ distribution at the 
Tevatron from ResBos and from two recent versions of {\tt PYTHIA}. The 
ResBos and {\tt PYTHIA} predictions have been normalized to the same area. 
The bottom figure shows the same in logarithmic scale.
} 
\label{fig:resbos_pythia_higgs_tev}
\end{figure}
A comparison of the two versions of {\tt PYTHIA} and of ResBos is also 
shown in Figure~\ref{fig:resbos_pythia_higgs_tev} for the case of Higgs 
production at the Tevatron with center-of-mass energy of 2.0 TeV (for a 
hypothetical SM Higgs mass of 100 GeV) .~\footnote{ As an exercise, events 
for an 80 GeV $W$ boson and an 80 GeV Higgs were generated at the 
Tevatron using {\tt PYTHIA}~5.7~\cite{mrennarun2}. A comparison of the 
distribution of values of $\hat{u}$ and the virtuality $Q$ for the two
processes indicates a greater tendency for the Higgs virtuality to be near
the maximum  value and for there to be a larger number of Higgs events
with positive $\hat{u}$.} 
The same qualitative features are observed as at 
the LHC: the newer version of {\tt PYTHIA} agrees better with ResBos in 
describing the low $p_T$ shape, and there is a falloff at high $p_T$ 
unless the larger virtuality is used for the parton showers. The 
default (rms) value of the non-perturbative $k_T$ (0.44 GeV)  was used for 
the {\tt PYTHIA} predictions for Higgs production. 

\section{Comparison with {\tt HERWIG}}

The variation between versions 5.7 and 6.1 of {\tt PYTHIA} gives an 
indication of the uncertainties due to the types of choices that can be 
made in Monte Carlos. The prescription that $\hat{u}$ be negative for all 
branchings is a choice rather than an absolute requirement.  Perhaps the 
better agreement of version 6.1 with ResBos is an indication that the 
adoption of the $\hat{u}$ restrictions was correct. Of course, there may 
be other changes to {\tt PYTHIA} which would also lead to better agreement 
with ResBos for this variable. 

\begin{figure}[t]
\begin{center}
\epsfxsize=12cm
\epsfysize=12cm
\vspace{-2cm}
\mbox{\epsfbox{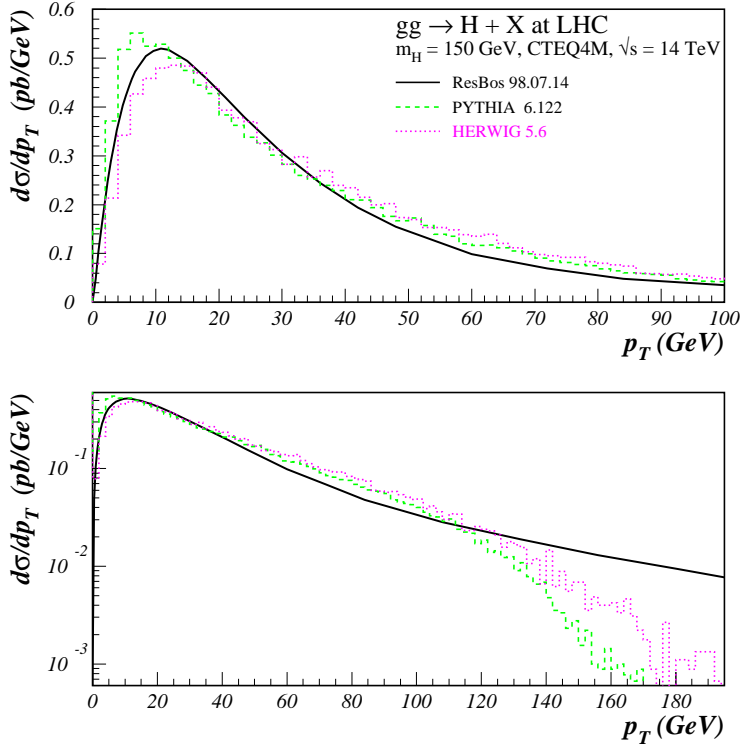}}
\end{center}
\caption{
\sf  A comparison of predictions for the Higgs $p_T$ distribution at the 
LHC from ResBos, a recent version of {\tt PYTHIA}, and {\tt HERWIG}. All 
predictions have been normalized to the same area.
The bottom figure shows the same in logarithmic scale.
} 
\label{fig:comparison_lhc}
\end{figure}
\begin{figure}[t]
\begin{center}
\epsfxsize=12cm
\epsfysize=12cm
\vspace{-2cm}
\mbox{\epsfbox{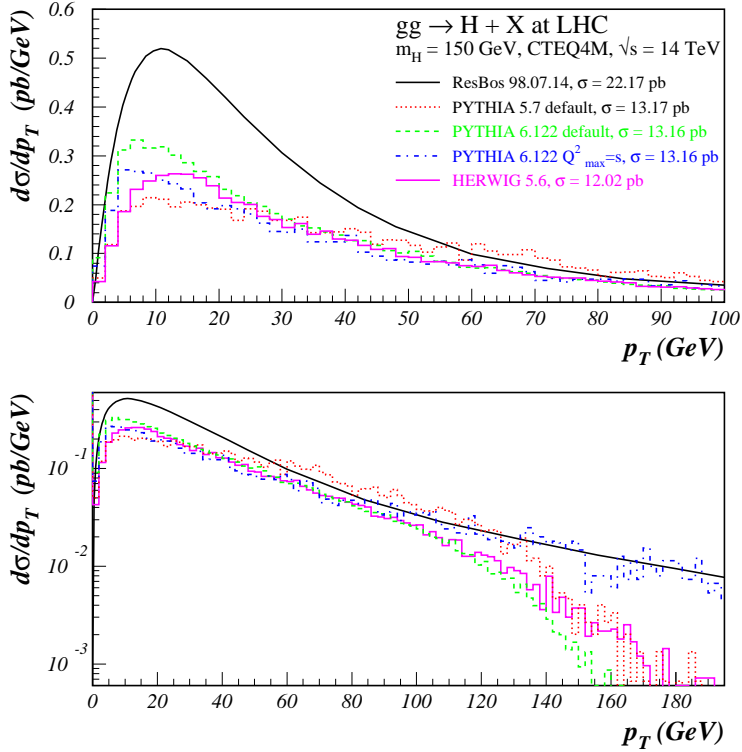}}
\end{center}
\caption{
\sf A comparison of predictions for the Higgs $p_T$ distribution at the 
LHC from ResBos, two recent versions of {\tt PYTHIA}, and {\tt HERWIG}. 
All predictions have their absolute normalizations. 
The bottom figure shows the same in logarithmic scale.
} 
\label{fig:comp_lhc_no_norm}
\end{figure}

Since there are a variety of choices that can be made in Monte Carlo 
implementations, it is instructive to compare the predictions for the 
$p_T$ distribution for Higgs production from ResBos and {\tt PYTHIA} with 
that from {\tt HERWIG} version 5.6. The {\tt HERWIG} prediction, 
for the Higgs $p_T$ distribution at the LHC, is shown in 
Figure~\ref{fig:comparison_lhc}, along with the {\tt PYTHIA} and ResBos 
predictions, all normalized to the ResBos prediction.~%
\footnote{The normalization factors (ResBos/Monte Carlo) are 
   1.68 for both versions of {\tt PYTHIA}, and 1.84 for {\tt HERWIG}.} 
(In all cases, the CTEQ4M parton distribution was used.) The predictions 
from {\tt HERWIG} and {\tt PYTHIA} 6.1 are very similar, with the {\tt 
HERWIG} prediction matching the ResBos shape somewhat better at low $p_T$. 
It is interesting that {\tt HERWIG} matches the ResBos prediction so closely
without the implementation of any kinematic cuts as in {\tt PYTHIA} 6.1. 
Perhaps the reason is related to the treatment of color coherence in the 
{\tt HERWIG} parton showering algorithm. 
For reference, the absolutely normalized predictions from ResBos,  {\tt 
PYTHIA}, and {\tt HERWIG} for the $p_T$ distribution of a 150 GeV Higgs at 
the LHC are shown in Figure~\ref{fig:comp_lhc_no_norm}.

\section{Non-perturbative $k_T$}

\begin{figure}[t]
\begin{center}
\epsfxsize=12cm
\epsfysize=12cm
\vspace{-2cm}
\mbox{\epsfbox{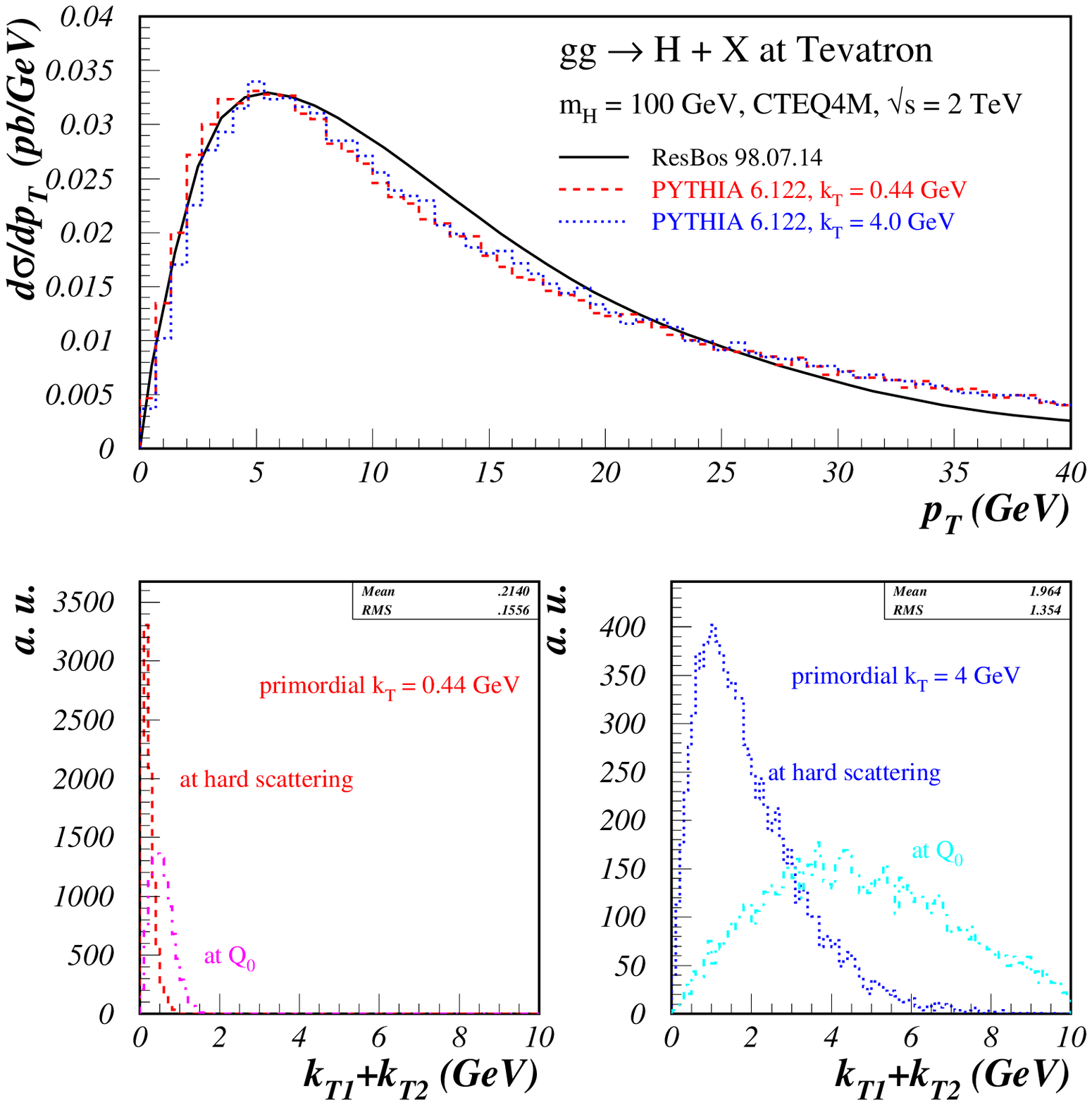}}
\end{center}
\caption{
\sf (top) A comparison of the {\tt PYTHIA} predictions for the $p_T$ 
distribution of a 100 GeV Higgs at the Tevatron using the default (rms) 
non-perturbative $k_T$ (0.44 GeV) and a larger value (4 GeV), at the 
initial scale $Q_0$ and at the hard scatter scale. Also shown is the 
ResBos prediction.
(bottom) The vector sum of the intrinsic $k_T$ ({\boldmath$k$}$_{T1}$+
{\boldmath$k$}$_{T2}$) for the two initial state partons at the initial 
scale $Q_0$ and at the hard scattering scale, for the two values of primordial 
$k_T$.
}
\label{fig:kt_higgs_tev}
\end{figure}

\begin{figure}[t]
\begin{center}
\epsfxsize=12cm
\epsfysize=12cm
\vspace{-2cm}
\mbox{\epsfbox{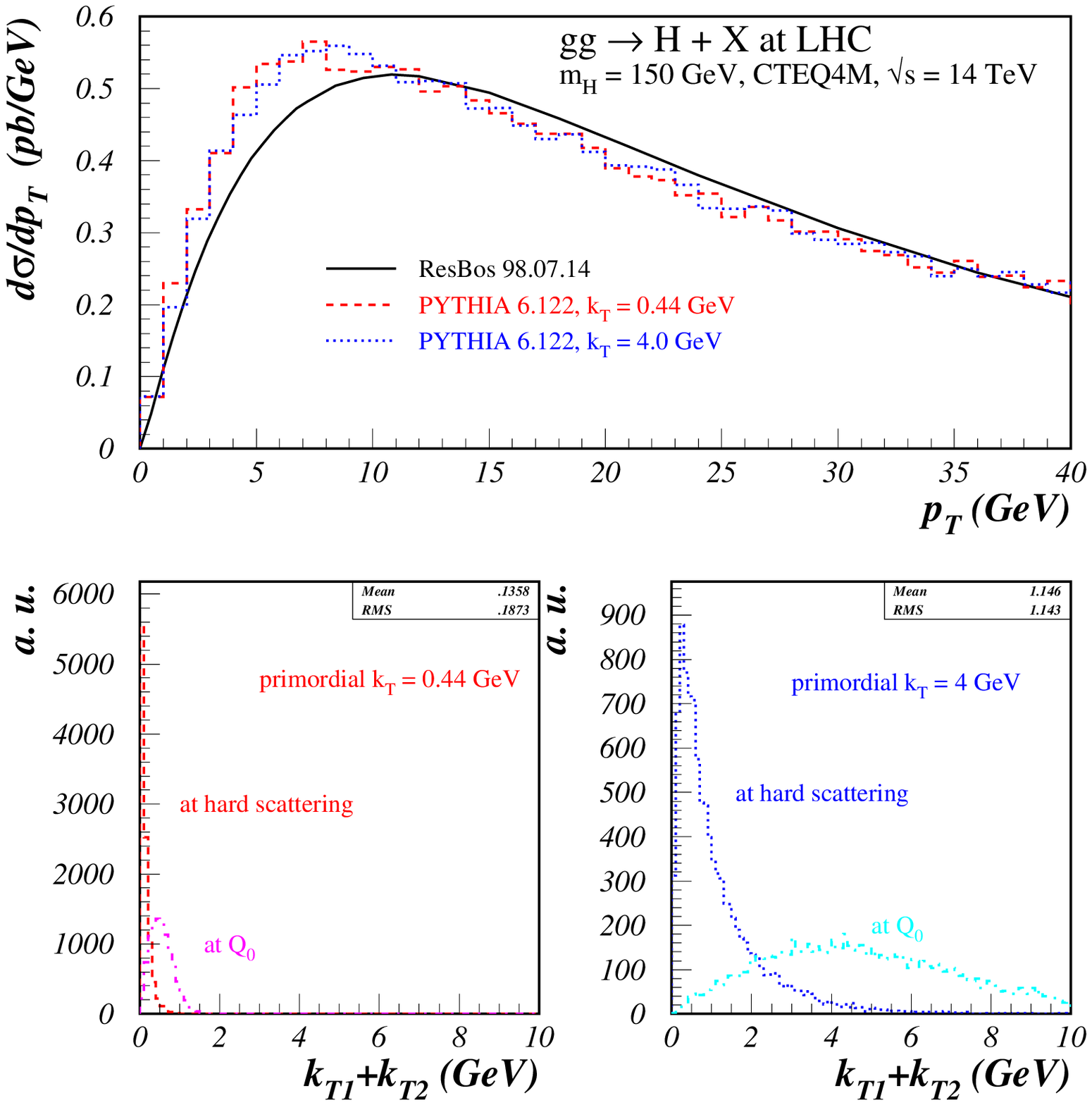}}
\end{center}
\caption{
\sf (top) A comparison of the {\tt PYTHIA} predictions for the $p_T$ 
distribution of a 150 GeV Higgs at the LHC using the default (rms) non-%
perturbative $k_T$ (0.44 GeV) and a larger value (4 GeV), at the initial 
scale $Q_0$ and at the hard scatter scale. Also shown is the ResBos 
prediction. (bottom) The vector sum of the intrinsic $k_T$ 
({\boldmath$k$}$_{T1}$+{\boldmath$k$}$_{T2}$) 
for the two initial state partons at the initial 
scale $Q_0$ and at the hard scattering scale, for the two values of 
primordial $k_T$.
}
\label{fig:kt_higgs_lhc}
\end{figure}

A question still remains as to the appropriate input value of non-perturbative 
$k_T$ in the Monte Carlos to achieve a better agreement in shape, 
both at the Tevatron and at the LHC.~\footnote{This has also been explored 
for direct photon production in Ref.~\cite{BaerReno}.}
In Figures~\ref{fig:kt_higgs_tev} 
and~\ref{fig:kt_higgs_lhc}  are shown  comparisons of ResBos and {\tt 
PYTHIA} predictions for the Higgs $p_T$ distribution at the Tevatron and the 
LHC. The {\tt PYTHIA} prediction (version 6.1 alone) is shown with 
several values of non-perturbative $k_T$. Surprisingly, no difference is 
observed between the predictions with the  different values of $k_T$, with 
the peak in {\tt PYTHIA} always being somewhat below that of ResBos. This 
insensitivity can be understood from the plots at the bottom of the two 
figures, which show the sum of the non-perturbative partonic initial state 
$k_T$'s ({\boldmath$k$}$_{T1}$+{\boldmath$k$}$_{T2}$) 
at $Q_0$ and at the hard scatter scale $Q$. Most of 
the $k_T$ is radiated away, with this effect being larger (as expected) 
at the LHC. The large gluon radiation probability from a $gg$ 
initial state and the greater phase space available at the LHC lead to a 
stronger degradation of the non-perturbative $k_T$ than was observed with 
$Z^0$ production at the Tevatron.

\begin{figure}[t]
\begin{center}
\epsfxsize=12cm
\epsfysize=12cm
\vspace{-2cm}
\mbox{\epsfbox{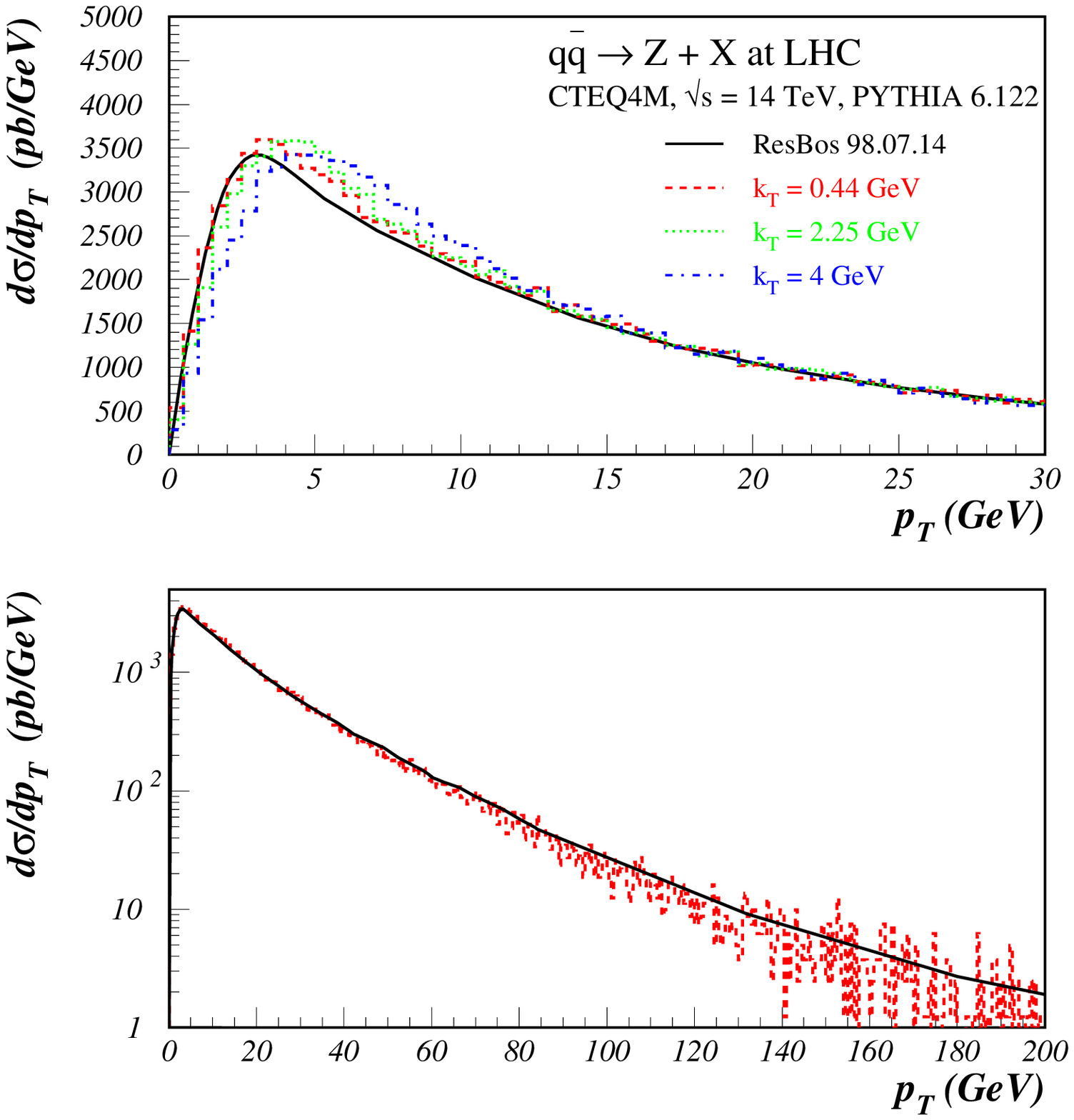}}
\end{center}
\caption{
\sf A comparison of the predictions for the $p_T$ distribution for $Z^0$ 
production at the LHC from {\tt PYTHIA} and ResBos, where several values 
of $k_T$ have been used to make the {\tt PYTHIA} predictions.
} 
\label{fig:z_lhc}
\end{figure}

\begin{figure}[t]
\begin{center}
\epsfxsize=12cm
\epsfysize=12cm
\vspace{-2cm}
\mbox{\epsfbox{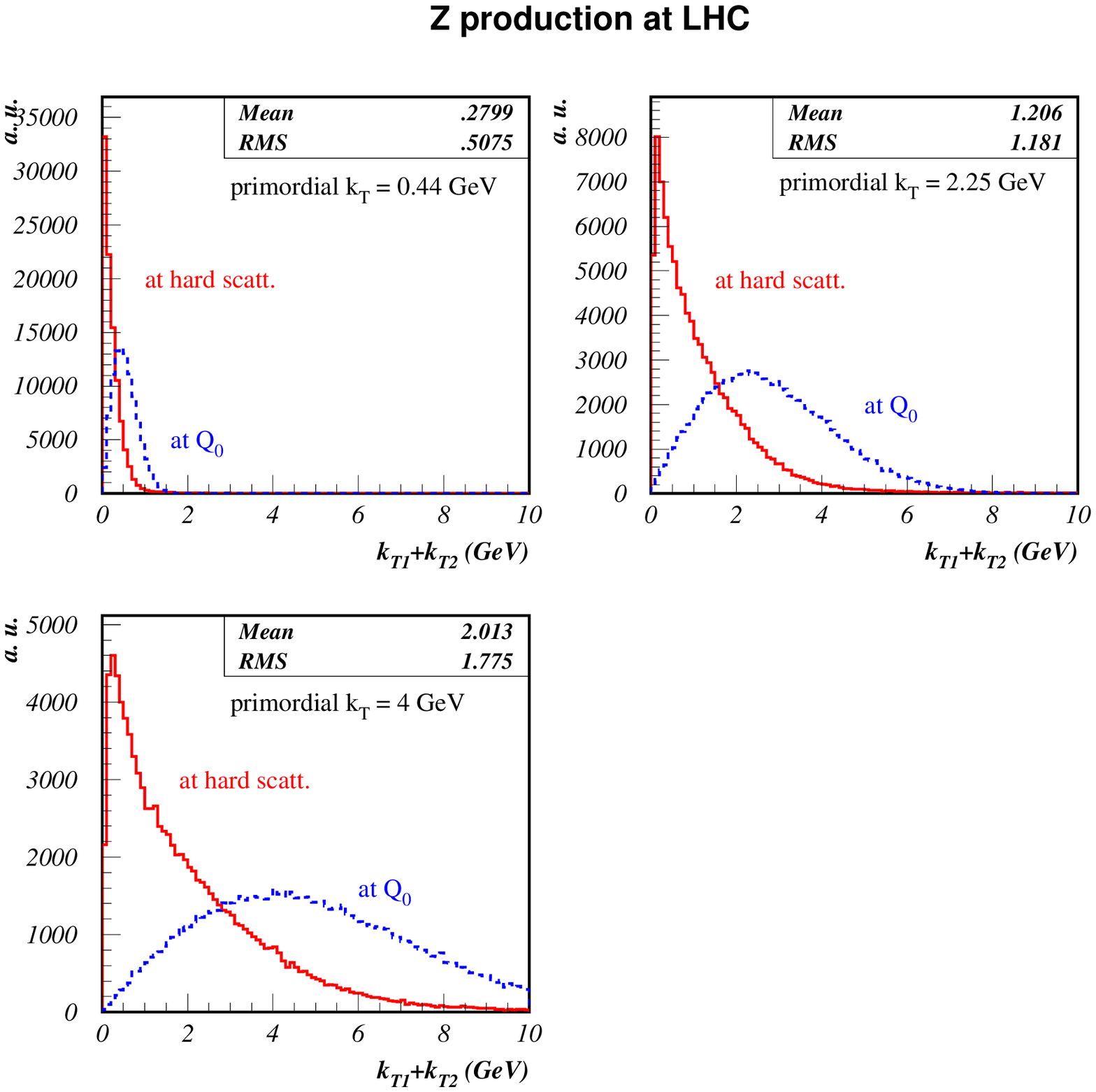}}
\end{center}
\caption{
\sf A comparison of the total initial state $k_T$
({\boldmath$k$}$_{T1}$+{\boldmath$k$}$_{T2}$)
distributions for $Z^0$ production at the
LHC from {\tt PYTHIA}, both at the initial scale $Q_0$ and at the hard 
scattering scale, for several (rms) values of the initial state $k_T$. 
The mean and rms numbers refer to the values at the hard scattering scale.
} 
\label{fig:z_lhc_kt}
\end{figure}

For completeness, a comparison of {\tt PYTHIA} and ResBos is shown in 
Figure~\ref{fig:z_lhc} for $Z^0$ boson production at the LHC. There are two 
points that are somewhat surprising. First, there is still a very strong 
sensitivity to the value of the non-perturbative $k_T$ used in the 
smearing. Second, the best agreement with ResBos is obtained with the default 
value (0.44 GeV), in contrast to the 2.15 GeV needed at the Tevatron 
(cf. Fig\ref{fig:run1_ee_pt}). Note 
again the agreement of {\tt PYTHIA} with ResBos at the highest values of 
$Z^0$ $p_T$ due to the explicit matrix element corrections applied. 
The sum of the incoming parton $k_T$ distributions, both at the scale 
$Q_0$ and at the hard scattering scale, are shown in 
Figure~\ref{fig:z_lhc_kt} for several different starting (rms) values of 
primordial $k_T$ (per parton). There is substantially less radiation for a 
$q\overline{q}$ initial state than for a $gg$ initial state (as in the case 
of the Higgs), leading to a noticeable dependence of the $Z^0$ $p_T$ 
distribution on the primordial $k_T$ distribution. 

\section{Conclusions}

An understanding of the signature for Higgs boson production at either the 
Tevatron or the LHC depends upon the understanding of the details of the 
multiple soft--gluon emission from the initial state partons. This 
soft--gluon radiation can be modeled either in a Monte Carlo or by an analytic 
resummation calculation, with various choices possible in both 
implementations. A comparison of the two approaches helps in understanding 
their strengths and weaknesses, and their reliability. The data from the 
Tevatron that either exist now, or will exist in Run 2, and from the LHC 
will be extremely useful to test both methods. 

\section{Acknowledgements}

We would like to thank Stefano Catani, Claude Charlot, Gennaro Corcella, 
Steve Mrenna, and Torbj\"orn Sj\"ostrand for useful conversations. We 
thank Willis Sakumoto for providing the figures for CDF $Z^0$ production, 
and Valeria Tano for HERWIG curves. C.B. and J.H. thank C.-P. Yuan for 
numerous discussions and blackboard lectures. This work was supported in 
part by DOE under grant DE-FG-03-94ER40833, and by NSF under grant 
PHY-9901946.

\vskip .5cm

At the final stage of preparing this manuscript, we became
aware of a new preprint \cite{Grazzini}, which studied the $A$ and $B$
functions for Higgs production up to $O(\alpha_S^4)$, and presented an
expression for the coefficient $B^{(2)}$.

\vskip 1cm



\begin{thebibliography}{999}

\bibitem{PoR2W}
Proceedings of {\it Physics at Run 2: Workshop on Supersymmetry and Higgs},
Summary Meeting, Batavia, IL, Nov. 19-21, 1998.


\bibitem{Abdullin} 
S.~Abdullin, M.~Dubinin, V.~Ilyin, D.~Kovalenko, V.~Savrin and 
N.~Stepanov, 
Phys.\ Lett.\  {\bf B431}, 410 (1998) [hep-ph/9805341].                  


\bibitem{BalazsNadolskySchmidtYuan} 
C.~Bal\'azs, P.~Nadolsky, C.~Schmidt and C.--P.~Yuan, 
hep-ph/9905551.


\bibitem{BalazsYuanZZ} 
C.~Bal\'azs and C.~P.~Yuan, 
Phys.\ Rev.\  {\bf D59}, 114007 (1999) [hep-ph/9810319].


\bibitem{BalazsYuanShortH} 
C.~Bal\'azs and C.--P.~Yuan, hep-ph/0001103.


\bibitem{Sjostrand:1985xi}
T.~Sj\"ostrand,
Phys.\ Lett.\ {\bf B157}, 321 (1985).


\bibitem{pythia} 
T.~Sj\"ostrand,
Comput.\ Phys.\ Commun.\ {\bf 82}, 74 (1994).


\bibitem{herwig} 
G. Marchesini, B.R. Webber, G. Abbiendi, I.G. Knowles, M.H. Seymour and 
L. Stanco, 
Comput.\ Phys.\ Commun.\ {\bf 67}, 465 (1992).


\bibitem{isajet} 
F.E. Paige, S.D. Protopescu, H. Baer and X. Tata, hep-ph/9810440. 


\bibitem{TDR} 
ATLAS Detector and Physics Performance Technical Design
Report, CERN/LHCC/99-14.


\bibitem{CSS}  
J.C.~Collins and D.E.~Soper, 
Phys.\ Rev.\ Lett.\ \textbf{48}, 655 (1982); 
%
Nucl.\ Phys.\ \textbf{B193}, 381 (1981), \textbf{B213}, 545(E)~(1983);
%
\textbf{B197}, 446 (1982); 

J.C. Collins, D.E. Soper and G. Sterman, 
{Nucl.~Phys.}~\textbf{B250}, 199~(1985).

For the application of the low $p_T$ factorization formalism to $Z^0$ boson and 
diphoton production see: 
\cite{BalazsYuanZZ,BalazsYuanWZ,BalazsBergerMrennaYuan}, and 

C.~Bal\'{a}zs, Ph.D. thesis, Michigan State University (1999), 
hep-ph/9906422. 


\bibitem{BCS}  
C. Bal\'azs, J.C.~Collins and D.E.~Soper, 
``Generalized factorization and resummation'',
in the proceedings of the {\it Workshop on Physics at TeV Colliders},
Les Houches, France, June 8-18, 1999.


\bibitem{BalazsYuanWZ}
C.~Bal\'azs and C.--P.~Yuan,
Phys.\ Rev.\ {\bf D56}, 5558 (1997) [hep-ph/9704258].


\bibitem{Brock}
F.~Landry, R.~Brock, G.~Ladinsky and C.--P.~Yuan,
hep-ph/9905391.


\bibitem{Yuan}
C.--P.~Yuan, Phys.\ Lett.\ {\bf B283}, 395 (1992).


\bibitem{BalazsMoriond}
C.~Bal\'azs, hep-ph/0008160.


\bibitem{carlschmidt} 
M. Grazzini, private communication. 
C. Schmidt, private communication. 


\bibitem{webber} 
S. Catani and B.R. Webber, 
Nucl.\ Phys.\ {\bf B349}, 635 (1991).


\bibitem{pythiacor} 
G. Miu and T. Sj\"ostrand, 
Phys.\ Lett.\ {\bf B449}, 313 (1999).


\bibitem{herwigcor} 
G. Corcella and M.H. Seymour, 
RAL-TR-1999-051 and hep-ph/9908388. 


\bibitem{isajetcor} 
H.~Baer and M.~H.~Reno,
Phys.\ Rev.\  {\bf D44}, 3375 (1991);
%
{\bf D45}, 1503 (1992).


\bibitem{mrenna} 
S. Mrenna, hep-ph/9902471.


\bibitem{Affolder:1999jh}
T.~Affolder {\it et al.}  [CDF Collaboration],
Phys.\ Rev.\ Lett.\ {\bf 84}, 845 (2000) [hep-ex/0001021].


\bibitem{corcella} 
G. Corcella, talk at the LHC workshop, October 1999.


\bibitem{cteq4} 
H.L. Lai, J. Huston, S. Kuhlmann, F. Olness, J. Owens, D. Soper, W.K. Tung, 
and H. Weerts, 
Phys.\ Rev.\ {\bf D55}, 1280 (1997).


\bibitem{sjostrandpc} T.~Sj\"ostrand, private communication. 


\bibitem{LY} 
G. Ladinsky, C.--P. Yuan, 
Phys.\ Rev.\ {\bf D50}, 4239 (1994).


\bibitem{cdfdiphot} 
F. Abe {\it et al.}, 
Phys.\ Rev.\ Lett.\ {\bf 70}, 2232 (1993); 

T. Takano, Ph.D. thesis, U. Tsukuba (1998); 

CDF Collaboration, paper in preparation.


\bibitem{d0diphot}
W. Chen, Ph.D. thesis, State University of New York, Stony Brook, NY (1997); 

D0 Collaboration, paper in preparation.


\bibitem{owens} 
P. Aurenche, A. Douri, R. Baier, and M. Fontannaz,
Z.\ Phys.\ {\bf C29}, 423 (1985);

B. Bailey, J. Owens, and J. Ohnemus, 
Phys.\ Rev.\ {\bf D46}, 2018 (1992); 

T. Binoth, J.P. Guillet, E. Pilon, and M. Werlen, 
hep-ph/9911340.


\bibitem{BalazsBergerMrennaYuan}
C.~Bal\'azs, E.L.~Berger, S.~Mrenna and C.--P.~Yuan,
Phys.\ Rev.\ {\bf D57}, 6934 (1998) [hep-ph/9712471];


\bibitem{pythiaman} 
{\tt PYTHIA} manual update for version 6.1,
{\tt http://www.thep.lu.se/$\sim$torbjorn/Pythia.html}.




\bibitem{corcellapc}
G. Corcella, private communication.


\bibitem{mrennarun2} 
S. Mrenna, talk at the {\it Physics at Run 2: Workshop on Supersymmetry 
and Higgs}, Summary Meeting, Batavia, IL, Nov. 19-21, 1998;

C.~Bal\'azs, J.~Huston, S.~Mrenna, and I.~Puljak, in the proceedings of
{\it Physics at Run 2: Workshop on Supersymmetry and Higgs},
Summary Meeting, Batavia, IL, Nov. 19-21, 1998. 


\bibitem{BaerReno}
H.~Baer and M.~H.~Reno,
Phys.\ Rev.\  {\bf D54}, 2017 (1996) [hep-ph/9603209].


\bibitem{Grazzini}
D.~de Florian and M.~Grazzini,
hep-ph/0008152.


\end{thebibliography}
\end{document}